\documentclass[aip,apl,reprint,amsmath,amsfonts,amssymb,floatfix]{revtex4-2}

\usepackage[T1]{fontenc}
\usepackage[utf8]{inputenc}
\usepackage[table]{xcolor}
\usepackage{graphicx}
\usepackage[final]{microtype}
\usepackage{subcaption}

\usepackage{dcolumn}
\usepackage{bm}
\usepackage{mathptmx}

\usepackage[colorlinks=true, linkcolor=blue, citecolor=blue, urlcolor=blue, bookmarks=true, breaklinks=true]{hyperref}
\usepackage{bookmark}

\newcommand{\joifrac}[2]{{#1}/{#2}}
\colorlet{grarEitt}{gray!8}
\colorlet{grarTvo}{gray!2}
\newcommand{\red}[1]{{\color{red}#1}}
\definecolor{darkgreen}{rgb}{0 0.6 0}

\begin{document}


\title{Space-Charge Limited Current from a Finite Emitter in Nano- and Microdiodes}


\author{Jóhannes Bergur Gunnarsson}
\affiliation{Department of Engineering, Reykjavik University, Menntavegur 1, IS-102 Reykjavik, Iceland}

\author{Kristinn Torfason}
\affiliation{Department of Engineering, Reykjavik University, Menntavegur 1, IS-102 Reykjavik, Iceland}

\author{Andrei Manolescu}
\affiliation{Department of Engineering, Reykjavik University, Menntavegur 1, IS-102 Reykjavik, Iceland}

\author{Ágúst Valfells}
\affiliation{Department of Engineering, Reykjavik University, Menntavegur 1, IS-102 Reykjavik, Iceland}


\begin{abstract}
We simulate numerically the classical charge dynamics in a microscopic, planar, vacuum diode with a finite emitter area and a finite number of electrons in the gap. We assume electrons are emitted under space-charge limited conditions with a fixed potential applied to the diode. The Coulomb interaction between all electrons is included using the method of molecular dynamics.  We compare our results to the conventional two-dimensional Child-Langmuir and explain how it is limited in applicability for sub-micron diameter emitters.  Finally, we offer some simple relations for understanding space-charge limited flow from very small emitters.   
\end{abstract}

\maketitle

\section{Introduction}

Space-charge limited current in diodes has been a subject of investigation for over a century~\cite{Zhang17a}. The classic Child-Langmuir law~\cite{Child11, Langmuir13} which describes the limiting current in a parallel-plate vacuum diode of infinite area is given by 
\begin{equation}
    J_{\rm CL} = \frac{4}{9}\varepsilon_0\sqrt{\frac{2q}{m}}\frac{V_g^{\joifrac{3}{2}}}{D^2}\, ,
    \label{eq:1DCL}
\end{equation}
where $\varepsilon_0$, is the permittivity of free space, $q$ represents the fundamental charge, $m$ the mass of an electron, $D$ the diode gap spacing and $V_g$ the potential across the diode.
The Child-Langmuir law has been extended to include different geometries~\cite{PhysRev.22.347, PhysRev.24.49, PhysRevLett.110.265007}, initial velocity~\cite{PhysRev.65.91}, relativistic~\cite{doi:10.1063/1.1657117}, and quantum effects~\cite{PhysRevLett.91.208303} among other things. An important modification to the Child-Langmuir law has to do with the situation where the emitting area is not infinite, but rather limited in dimension~\cite{PhysRevLett.87.278301, doi:10.1063/1.3243474, PhysRevLett.77.4668, PhysRevLett.87.145002, doi:10.1063/1.1913612}. This has obvious importance, as in a great number of practical devices the length scale of the emitting area is equal to or less than other length scales characterising the diode. An elegant theoretical derivation of the two-dimensional Child-Langmuir law is to be found in Reference~\cite{PhysRevLett.87.278301} and extended to a greater number of emitter shapes in~\cite{doi:10.1063/1.1913612}. The general form of the two-dimensional Child-Langmuir law is
\begin{equation}
    J_{\rm 2D\,CL} = J_{\rm CL}(1+G)\, ,
    \label{eq:2DCL}
\end{equation}
where $J_{\rm CL}$ is the Child-Langmuir current density as given by \autoref{eq:1DCL} and $G$ is a geometrical correction factor dependant on the shape of the emitting area, it's scale length, and the diode gap.

Important assumptions in the theoretical treatment are that the current density is continuous and uniform throughout the emitter area and that the beam does not expand transversely as it propagates across the diode gap. Although it is well known that the current density is generally higher at the emitter edge than in the interior region ~\cite{PhysRevLett.87.145002}, the assumption of uniform current density is generally valid for emitters of macroscopic length scales. At microscales one may anticipate that the wingtip structure of the current density profile will begin to play an important role. As will transverse expansion and discrete particle effects. In this paper molecular dynamics based simulations will be used to investigate how the space-charge limited current deviates from the theoretical estimate. The molecular dynamic approach is particularly well suited for this work as at the length scales and current densities involved it is anticipated that discrete particle effects will become important.

\section{Model and theoretical background}
The system under consideration consists of an infinite anode and cathode with a gap spacing $D$, and applied gap voltage $V_g$. Emission from the cathode is restricted to a circular area of radius, $R$. Electrons are emitted with negligible emission velocity. Electron emission and propagation is calculated using the same molecular dynamics approach for simulating space-charge limited dynamics as in previous papers from our research group~\cite{PhysRevLett.104.175002, doi:10.1063/1.4914855, kt_inhomo, Ilkov2015}. In short, the algorithm is the following: An emission site is randomly selected on the emitting area of the cathode. If the electric field at that site is oriented such that it would accelerate and electron away from the cathode, an electron is placed $1\ \mathrm{nm}$ above the cathode surface thus affecting the overall electric field. If the electric field at the site is not favorable for acceleration, no electron is placed at that location and a failure to place is recorded. This process is repeated until $100$ sequential failures to place have been recorded, indicating that there is no place on the emitting part of the cathode surface that has a favorably oriented field, hence the space-charge limit has set in. At this point direct Coulomb interaction is used to calculate the net force acting on every electron in the vacuum gap, the time-step in the simulation is advanced, and the Velocity-Verlet method is used to calculate the new positions and velocities of the electrons. Any electrons that exceed the boundaries of the diode are removed from the system. The simulation progresses through this procedure of electron placement/emission and advancement for any number of time steps. This method of electron injection ensures a self-consistent space-charge limited current density across the emitter. The current is calculated by use of the Ramo-Shockley theorem~\cite{doi:10.1063/1.1710367, 1686997}. The equation is
\begin{equation}
    I = \frac{q}{D}\sum_i v_{z,\, i}\, ,
\end{equation}
where \(q\) is the electron charge, \(v_z\) the component of the instantaneous velocity that is normal to the cathode surface, and \(D\) the gap spacing.

For the simulations the applied electric field (which is the electric field in the absence of space-charge) is kept constant; for most runs at a value of 
$V_g/D =1\ \mathrm{V}/\mu{\mathrm{m}}= 1\ \mathrm{MV/m} $, but we also conduct runs at other fixed values of the applied field.
The parameter, $\alpha$, hereafter called the aspect ratio is defined as
\begin{equation}
    \alpha = \frac{D}{R}\, ,
\end{equation}
where $R$ is the radius of the emitting area. The results described in this paper come from 91 sets of simulations. For the 61 sets with an applied field of $1\ \mathrm{MV/m}$ we run five different values of the aspect ratio: $\alpha = 2,\ 5,\ 10,\ 20,\ \textnormal{and}\ 50$. 50 of these simulations correspond to $V_g = 1,\dots,10$ for each aspect ratio. 
We add 11 extra sets for $\alpha = 20$ and $\alpha = 50$ to extend the range of the data. For $\alpha = 20$, these extra sets are for the following voltages: $V_g = 0.02,\ 0.05,\ 0.1,\ 0.2\ \textnormal{and } 0.5\ [{\rm V}]$, corresponding to a radius of $R =  1,\ 2.5,\ 5,\ 10 \ \textnormal{and } 25\ [{\rm nm}] $ respectively. 
For the aspect ratio of $\alpha = 50$ the extra sets are for $V_g = 0.05,\ 0.1,\ 0.2,\ 0.5,\ 20\ \textnormal{and } 40\ [V]$, corresponding to a radius of $R = 1,\ 2,\ 4,\ 10,\ 20,\ 400\ \textnormal{and } 800\ [{\rm nm}] $. Data for these 11 additional sets is only shown in \autoref{fig:R_vs_H_all_alpha} and \autoref{fig:current_vs_radius_all_asymptotes}. Additionally, we ran 30 sets for $\alpha = 10$ with fixed field values of $2\ \mathrm{MV/m}$, $5\ \mathrm{MV/m}$, and $10\ \mathrm{MV/m}$ and radius values of $R = 0.1,\dots,1\ \mathrm{\mu m}$. The are shown in \autoref{fig:different_E_H}.


From the generalized Child-Langmuir treatment of Koh et~al.~\cite{doi:10.1063/1.1913612} one expects the geometrical correction of \autoref{eq:2DCL} to be $G=D/(4R)$, since the emitting area is circular.
From the assumption that the current density is uniform across the emitter one expects the current to be 
\begin{equation}
    I_{\rm 2D} = \pi R^2 J_{\rm 2DCL} =\frac{\pi}{9}\varepsilon_0\sqrt{\frac{2q}{m}}V_g^{\joifrac{3}{2}}\left(\frac{4}{\alpha^2}+\frac{1}{\alpha} \right)\, .
    \label{eq:I_2D}
\end{equation}
In subsequent discussion of the simulation results, \autoref{eq:I_2D} will be used as the theoretical comparison.

\section{Results and analysis}
\begin{figure}
    \includegraphics[width = 0.95\linewidth]{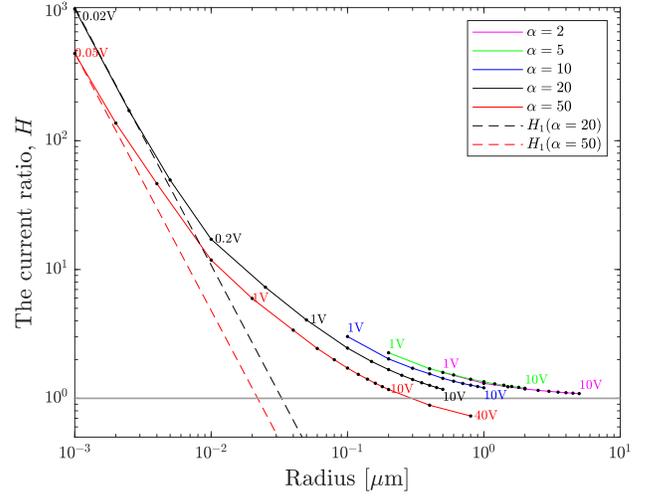}
    \caption{Ratio $H$, of the current  calculated from the simulation, to the current calculated from \autoref{eq:I_2D}. The dashed lines show the current ratio, for different aspect ratios, in the Coulomb blockade limit as described by \autoref{eq:H_1}.
    }
    \label{fig:R_vs_H_all_alpha}
\end{figure}

We begin by examining the ratio of the current calculated from the simulation, $I_{\rm sim}$, and the current predicted by \autoref{eq:I_2D}. For brevity we define the ratio as $H = I_{\rm sim}/I_{\rm 2D}$ in \autoref{fig:R_vs_H_all_alpha}.
It is immediately apparent that \autoref{eq:I_2D} is generally not applicable for radii less than 10 $\mu{\mathrm{m}}$, at least for the given applied field of $1\ \mathrm{MV/m}$. For "large" radii, and smaller aspect ratios, \autoref{eq:I_2D} gives a good approximation to the true value of the current. The reasons for this are most likely that edge emission is of greater importance for small radii, transverse beam expansion is of greater relative importance for small diameter emitters, as are discrete particle effects. In addition, it may be that transverse expansion of the beam plays an important role for beams with large aspect ratios.

\begin{figure}
    \includegraphics[width = 0.95\linewidth]{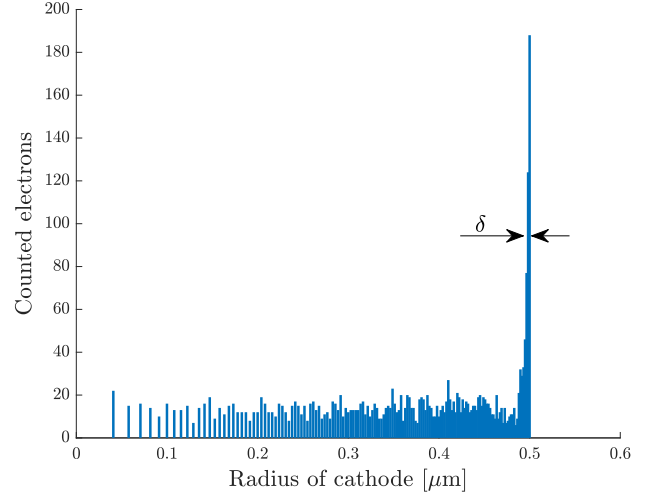}
    \caption{Histogram of current density as a function of radius,~$r$, at the cathode surface. $\alpha = 10$, $V_g = 5\ V$}
    \label{fig:radius_vs_electrons}
\end{figure}

An examination of the current density as a function of radius confirms that the emission from the edge can be considerably higher than from the bulk region as can be seen in \autoref{fig:radius_vs_electrons}. Note that in the bulk the current is quite uniform as it is assumed to be in the theoretical treatment leading to \autoref{eq:I_2D}. We define the edge region as an annulus covering the area of higher current density. It constitutes the part of the emitter where the radius, $r$, lies in the interval $R - \delta < r < R$, with $\delta$ defined in the following manner. Let $R_{1/2}$ denote the radius where the current density reaches half of its maximum value. Then we pick $\delta$ such that an annulus with outer radius $R$ and inner radius $R_{1/2}$, has the same area as an annulus with outer radius of $R_{1/2}$ and inner radius of $R-\delta$. Using this definition, and recording the point of origin of all electrons in the simulation, we can calculate the fraction of electrons that are emitted from the edge in a consistent manner.

\begin{figure}
    \includegraphics[width = 0.95\linewidth]{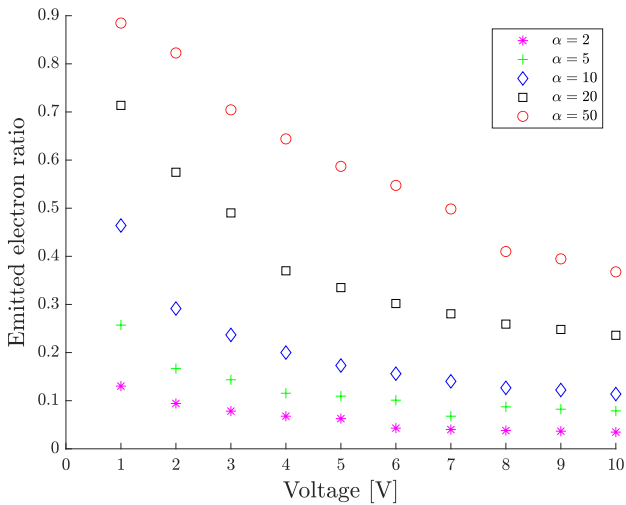}
    \caption{Fraction of the charge originating in the edge region of the emitting area.}
    \label{fig:emitted_ratio}
\end{figure}

\autoref{fig:emitted_ratio} shows how the fraction of charge coming from the edge varies with the applied voltage for the different aspect ratio values. Equivalently, since with our constant electric field $V_g [\rm V]=E_0[{\rm \joifrac{MV}{m}}]\ \alpha \ R [\mathrm{\mu m}]$, the emitted electron ratio can be evaluated as function of the emitter radius. We note that for small radii the fraction of charge coming from the edge is considerable, and in fact may constitute the bulk of the current coming from the emitter. 
Of course, even if the current density were uniform across the entire emitter, we would anticipate that the fraction of the current coming from an edge region of width $\delta$ would scale as $1/R$ with emitter radius, due to the fact that the area of the edge region divided by the total area of the circular patch is equal to $2\delta/R$. However, as the current density of the edge area is considerably higher than that of the bulk area we observe that the scaling with radius is quite different from $1/R$.

Now, consider a diode where electrons are emitted from a single point at the center of the emitting area of the cathode, and the diode gap is small enough that only one electron can be in the gap at a time. The maximum gap size such that only one electron can be accommodated at a time is found by using the Coulomb blockade condition that an electron cannot be released from the cathode until the retarding field due to the previously released electron no longer exceeds the applied field. This means that for an applied field, $E_0$, the maximum gap size allowed to accommodate only a single electron from a point emitter is $D_{\rm max} = \sqrt{q/(2\pi \varepsilon_0 E_0)}$ (taking into account both the electron and its image charge), which corresponds to \(54\ \mathrm{nm}\) for an applied field of $1\ \mathrm{MV/m}$.  The average current for a system of this type is $\langle I \rangle = q/\tau$, where $\tau = D\sqrt{2m/(qV_g)}$  is the time for an electron to transit the diode gap. Using this definition we may calculate the ratio, between the average current for a single electron, and the theoretical value derived from the continuous model of \autoref{eq:I_2D}, as 
\begin{equation}
    H_1 = \frac{\langle I \rangle}{I_{\rm 2D}}=\frac{9q}{2\pi \varepsilon_0 E_0(4+\alpha)R^2}\, .
    \label{eq:H_1}
\end{equation}

Referring back to \autoref{fig:radius_vs_electrons} we show how the values of $H$ from simulation approach $H_1$ asymptotically for different aspect ratio values. The point emitter approximation becomes ever more accurate as the emitter radius decreases. For small radii and large aspect ratios the single electron assumption is met.

\begin{figure}
\begin{subfigure}{1.0\linewidth}
    \caption{}
    \includegraphics[width = 0.95\linewidth]{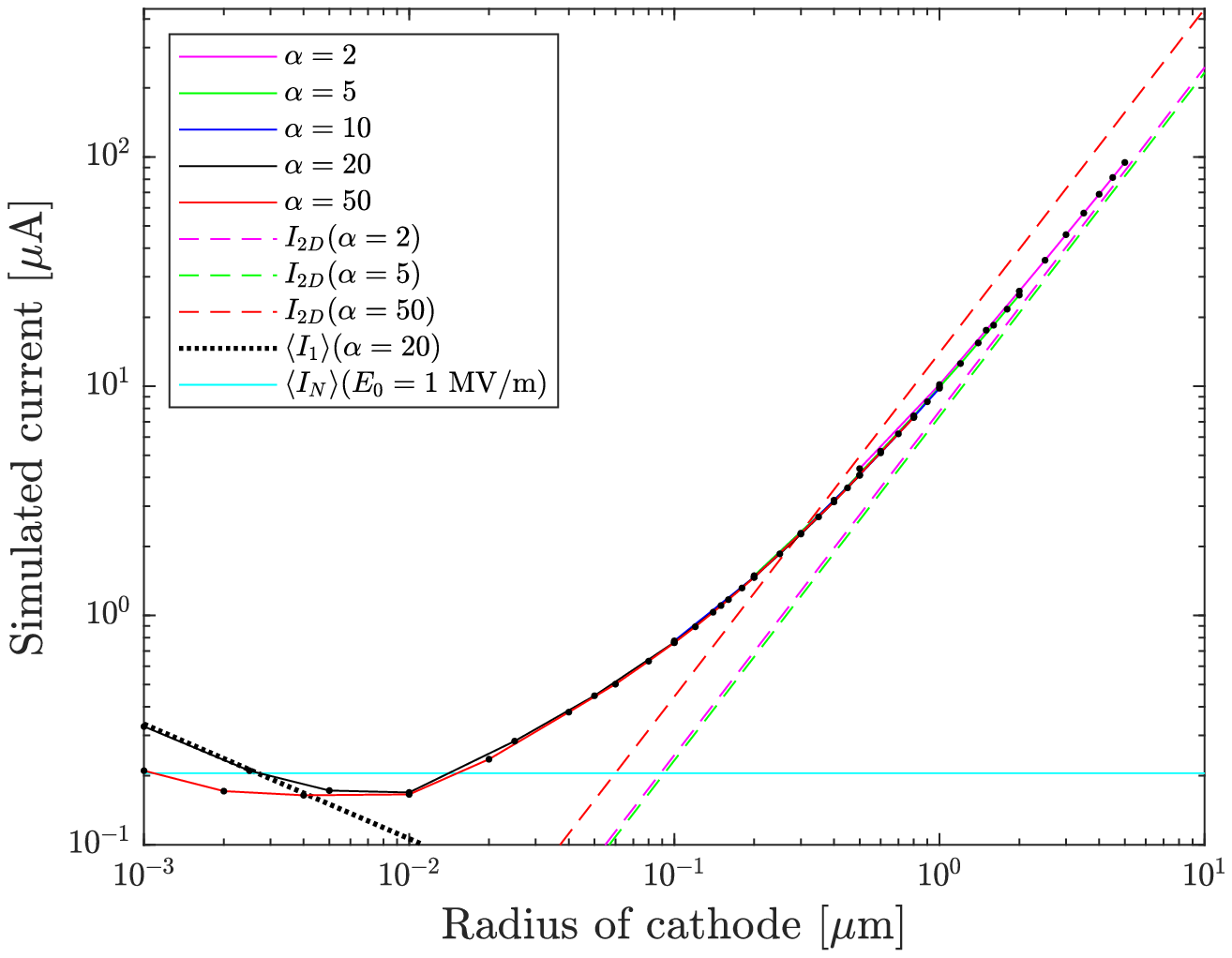}
    \label{fig:current_vs_radius_all_asymptotes}
\end{subfigure}
\begin{subfigure}{1.0\linewidth}
    \caption{}
    \includegraphics[width = 0.95\linewidth]{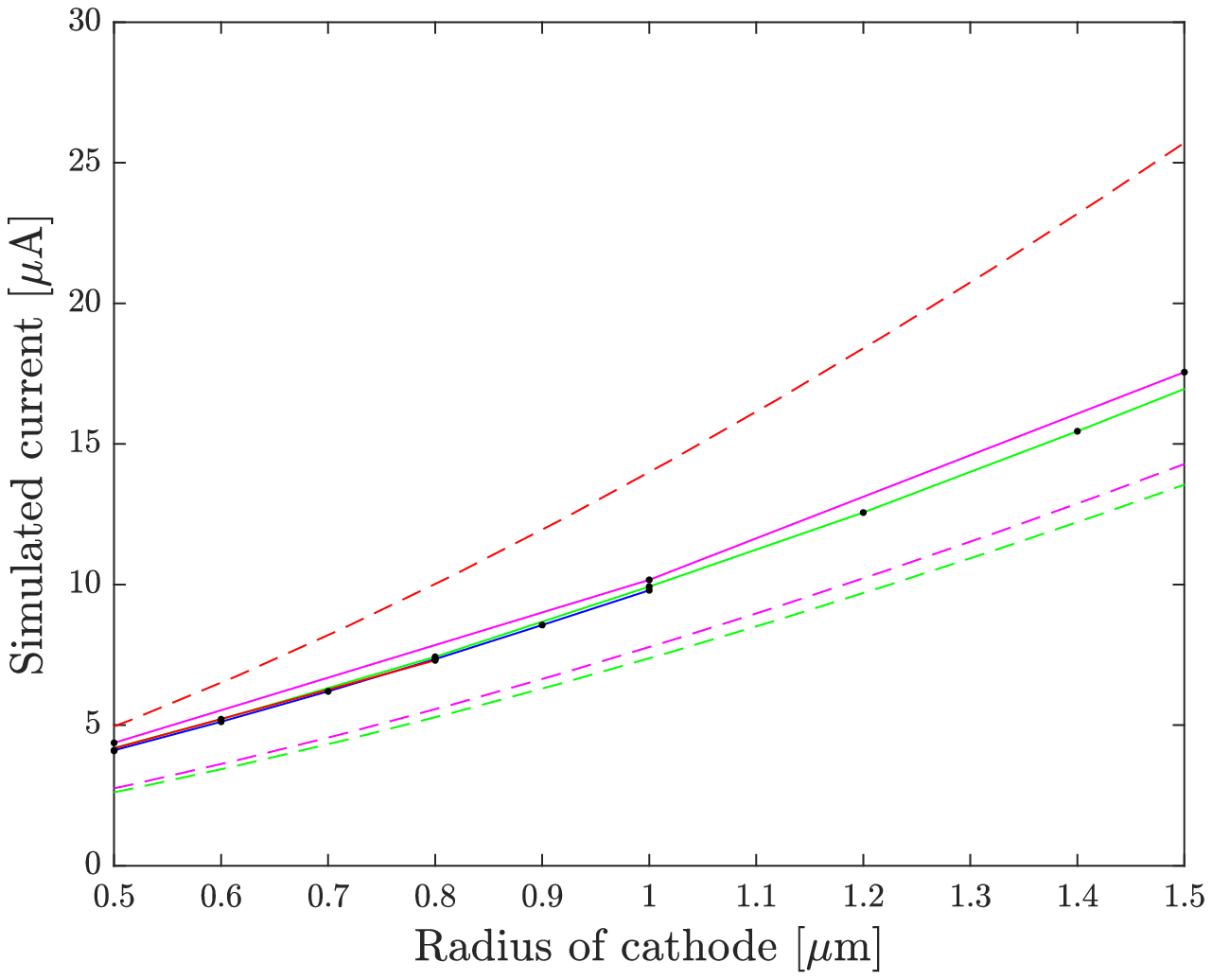}
    \label{fig:current_vs_radius_all_asymptotes_zoomed}
\end{subfigure}

\caption{ (a) Current versus emitter radius. Solid lines show simulation results. Dotted line shows the current as predicted by the single electron model of \autoref{eq:I_single_electron}, while the dashed lines show the current according to \autoref{eq:I2D}, for different aspect ratios. Horizontal line shows current from \autoref{eq:IN} (b)  Detail from (a) in linear scale. Note that the simulated current for $\alpha = 50$ (red line) coincides with the current for other aspect ratios rather than following the expected asymptote for $\alpha = 50$ (red dashed).}
\end{figure}

Let us further consider the total current extracted from the cathode. By using the relation $V_g = \alpha E_0 R$ we may recast \autoref{eq:I_2D} as
\begin{equation}
    I_{\rm 2D} = \frac{\pi}{9}\varepsilon_0 \sqrt{\frac{2q}{m}}E_0^{\joifrac{3}{2}}\left(\frac{4}{\sqrt{\alpha}}+\sqrt{\alpha}\right)R^{\joifrac{3}{2}}\, ,
    \label{eq:I2D}
\end{equation}
so that for a fixed applied field and aspect ratio the current scales as $R^{\joifrac{3}{2}}$. \autoref{fig:current_vs_radius_all_asymptotes}. depicts simulation results for current versus emitter radius for a constant value of applied field and five different fixed values of the aspect ratio, $\alpha$. Also shown are the predicted curves derived from \autoref{eq:I2D}. We see that, asymptotically, the current scales as $R^{\joifrac{3}{2}}$ with increasing radius, as is predicted by \autoref{eq:I2D}. However, We also note that when the radius is greater than $100\ \mathrm{nm}$ the simulated current is seemingly independent of the aspect ratio, unlike the current predicted by \autoref{eq:I2D}. As can be seen in more detail in \autoref{fig:current_vs_radius_all_asymptotes_zoomed} the predicted asymptotes for $\alpha = 2$ and $\alpha = 5$ are very similar, and it is toward this asymptote that all of the simulated currents tend, even for larger aspect ratios. A speculative explanation follows. For two diodes with the same applied field and emitter radius, but different aspect ratios, the gap spacing will differ in proportion to the aspect ratios. Thus, a beam with an aspect ratio of 50 will propagate across a gap that is ten times as long as does the beam with an aspect ratio of 5. According to \autoref{eq:I2D} the former beam should also have a considerably higher current density and thus the transverse space-charge force near the cathode is also greater. This means that the beam with the higher aspect ratio will experience a larger transverse force and have a longer transit time, so that we may expect it to expand considerably as it travels even a short distance from the cathode. This could result of giving it the appearance of having an effectively larger emission area which would result in an "effectively smaller" aspect ratio. We have not run simulations to test this hypothesis but they may be included in future work.  

We now turn our attention to the region where the radius is less than $10\ \mathrm{nm}$. Here we see that the current increases with decreasing emitter radius, and the aspect ratio affects the current. The reason for this is straightforward if one considers the point emitter model. When the gap spacing is less than $54\ \mathrm{nm}$ (recall that  $D=\alpha R$), there is only one electron in the gap and the average current is given by
\begin{equation}
    \langle I_1 \rangle = \frac{q}{\tau} = q\sqrt{\frac{qE_0}{2m\alpha R}}\, ,
    \label{eq:I_single_electron}
\end{equation}
which explains why a smaller aspect ratio and radius lead to a larger average current. Of course, there are limits to the applicability of this classical model. As the gap spacing is further reduced, quantum effects must be taken into account~\cite{PhysRevLett.91.208303} but that is beyond the scope of this paper.

Next we look at the current from a point emitter where the gap size is large enough that it can accomodate $N > 1 $ electrons. As an approximation we may make the assumption that electrons emitted from the point source do not interact, and that the period between electrons being released from the cathode is equal to the time that it takes for a single electron to reach an elevation above the cathode of $z_*=\sqrt{q/(2\pi \varepsilon_0 E_0)}$. This time is equal to $\tau_* = \sqrt{2mz_*/(qE_0)}$, which leads to an average current of 
\begin{equation}
    \langle I_N \rangle = \frac{q}{\tau_*} = \left(\frac{q^5\pi\varepsilon_0}{2m^2}\right)^{\joifrac{1}{4}}E^{\joifrac{3}{4}}\, ,
    \label{eq:IN}
\end{equation}
which is independent of the gap spacing and aspect ratio. The approximations made to obtain \autoref{eq:IN} result in a slight overestimation of the current, since it will not only be the electron nearest to the cathode that contributes to the Coulomb blockade, and thus it will have to travel further away from the point of emission before another electron can be released. \red{\autoref{fig:current_vs_radius_all_asymptotes}} depicts $\langle I_N \rangle$ in relation to the current values from simulations. As can be seen it is slightly above the minimum value of the current curve.

\begin{figure}
\begin{subfigure}{1.0\linewidth}
    \caption{}
    \includegraphics[width = 0.95\linewidth]{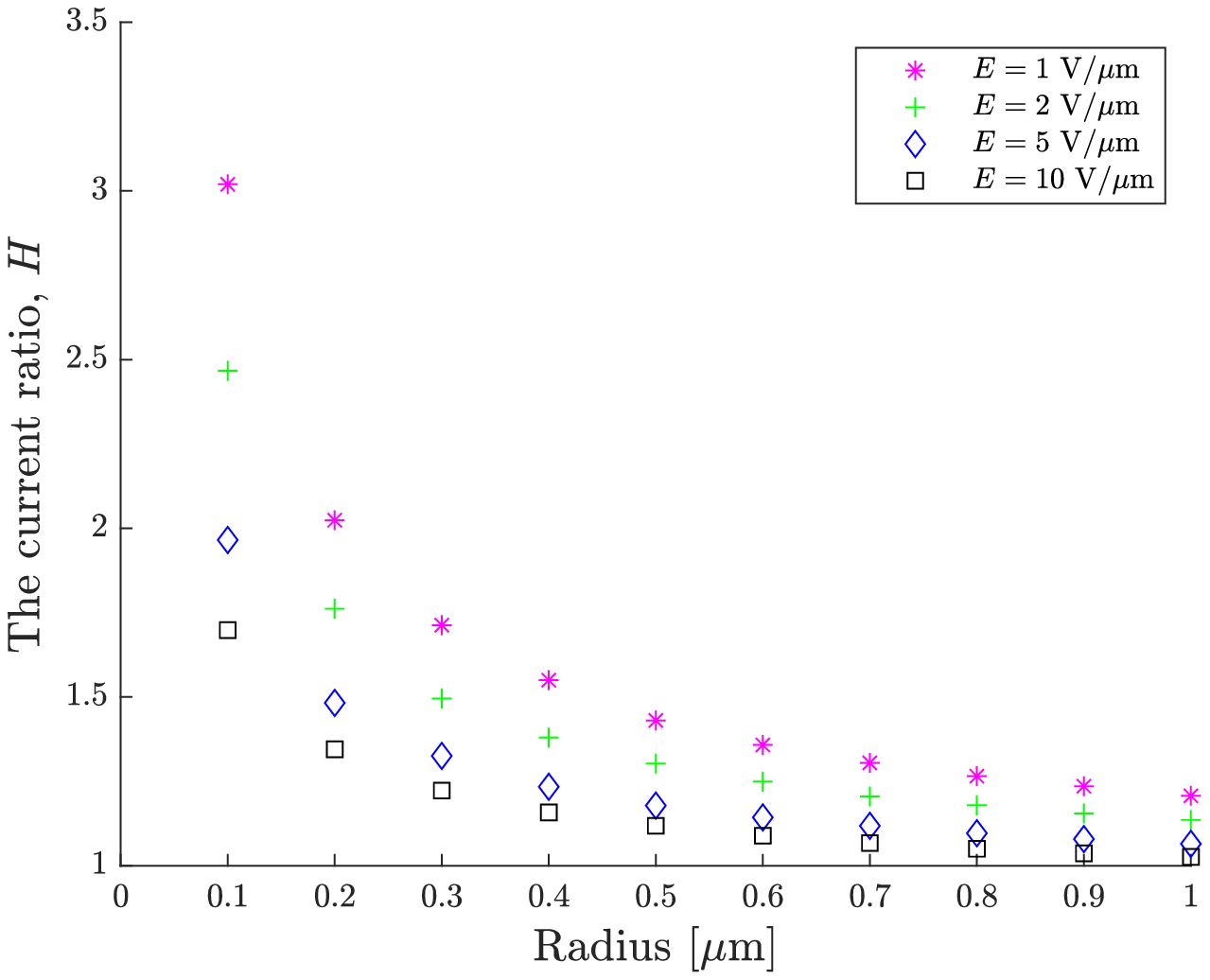}
    \label{fig:different_E_H}
\end{subfigure}\linebreak
\begin{subfigure}{1.0\linewidth}
    \caption{}
    \includegraphics[width = 0.95\linewidth]{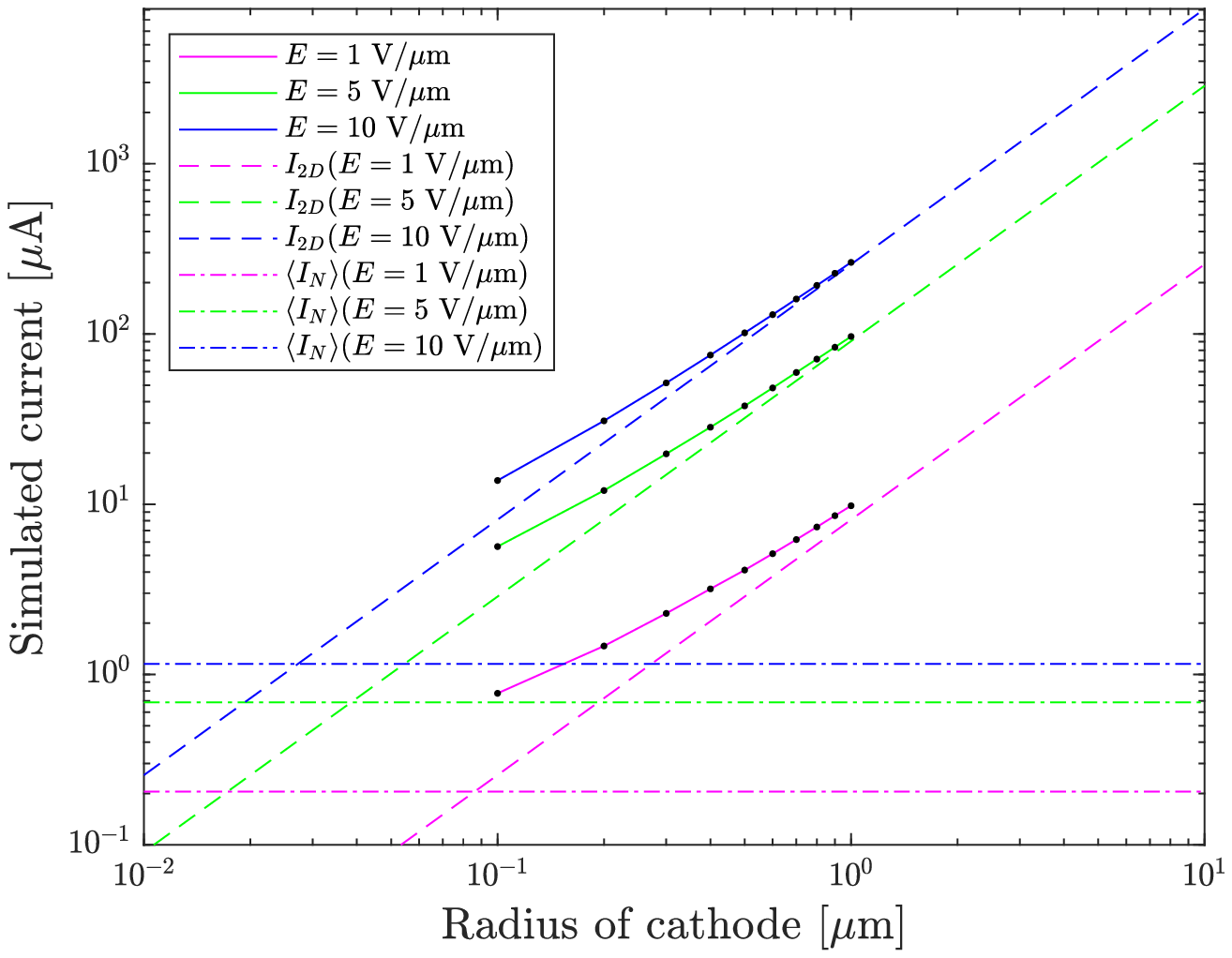}
    \label{fig:different_E_I}
\end{subfigure}
\caption{Effects of varying the applied field for a fixed aspect ratio $\alpha = 10$. (a) The current ratio, $H$ as a function of field. (b) The simulated current and expected asymptotes for select values of applied field and $\alpha = 10$. obtained from \autoref{eq:I2D} and \autoref{eq:IN}}
\end{figure}

At last, we examine the effect of the applied electric field on deviation from the conventional two-dimensional Child-Langmuir law. 
\autoref{fig:different_E_H} shows the ratio, $H$, for an aspect ratio of $\alpha = 10$, and four different values of the applied field strength. 
We see that, for a given emitter radius, \autoref{eq:I2D} gives a more accurate estimate of the actual current as applied field strength increases. 
This can be understood from a couple of considerations. First of all, one sees that the asymptote predicted by \autoref{eq:I2D} is shifted to smaller values of radius, in particular the point of intersection between the current predicted by \autoref{eq:IN} and \autoref{eq:I2D} shifts to a lower value of radius. 
This is the radius where the point emitter model becomes more appropriate and this can clearly be seen in \autoref{fig:different_E_I}. Secondly, if one considers the maximum transverse distance, $\Delta r$, over which an electron can effectively influence the surface electric field to a degree of $\Delta E$ is $\Delta r = \sqrt{q/(3 \sqrt{3} \pi \varepsilon_0\Delta E)}$, then the spacing between electrons emitted from the cathode will be roughly $\Delta r$. This means electron emission points on the cathode can be more closely spaced with higher applied fields, extending the continuous model to smaller radii.

\section{Summary and conclusion}
In this work we examine the physics of space-charge limited emission from a circular emitter of microscopic radius in a planar system. We establish that, for aspect ratios of 5 or smaller, and a radius greater than $10\ \mathrm{\mu m}$, the conventional theory for the two-dimensional Child-Langmuir law is sufficiently accurate. However, the conventional theory can be extended to smaller radii as the applied field strength increases. Our results also show, that for larger values of the aspect ratio, has the same asymptotic behavior, at large radii, as that for an aspect ratio of 2. We show that for small emitter radius and sufficiently large aspect ratio, a point emitter model gives a reasonably accurate estimate for the current that scales as $E_0^{\joifrac{3}{4}}$ but is independent of aspect ratio and emitter radius. Finally, for small emitter radius and small gap spacing, such that the diode can only accommodate one electron, the current scales with the gap spacing as $D^{-\joifrac{1}{2}}$ and with the applied field as $E_0^{\joifrac{1}{2}}$.

These results, are useful as they help establish the limitations of applicability of the conventional two-dimensional Child-Langmuir law, and extend our understanding of space-charge limited emission to finite emitter areas of sub-micron length scale. 

Applications may range from modelling emission from adsorbates on a cathode surface to emission from the apex of a field emitter. We have modelled a planar system with a uniform vacuum electric field, which is likely not applicable to a sharp field emitter where the electric field varies rapidly in the vicinity of the field emitter. Work on that problem is underway.

\begin{acknowledgments}
This material is based upon work supported by the Air Force Office of Scientific Research under award number FA9550-18-1-7011, and by the Icelandic Research Fund grant number 174127-051.
Any opinions, findings, and conclusions or recommendations expressed in this material are those of the authors and do not necessarily reflect the views of the United States Air Force.
\end{acknowledgments}

\bibliography{bibliography}
\end{document}